\documentclass[aip,reprint,groupedaddress]{revtex4-1}

\usepackage{graphicx}
\usepackage{amssymb}
\usepackage[fleqn]{amsmath}
\usepackage[main=english]{babel}
\usepackage[colorlinks=true,bookmarks=false,citecolor=blue,urlcolor=blue]{hyperref}

\begin{document}


\title{Leveraging natural fluctuations for matrix-based aberration correction in photoacoustic imaging} 



\author{Yevgeny Slobodkin}
\author{Ori Katz}
\email[]{orik@mail.huji.ac.il}
\affiliation{\mbox{Institute of Applied Physics, The Hebrew University of Jerusalem, Jerusalem 9190401, Israel}}


\date{\today}

\begin{abstract}
Photoacoustic imaging is the leading technique for deep tissue optical imaging, allowing single-shot imaging at depths. However, its resolution may be limited by acoustic aberrations, caused by natural unknown heterogeneities in the tissue speed of sound. 
In recent years, reflection-matrix based scattering-compensation techniques have been successfully employed in ultrasound, optics, and seismology, to computationally correct such distortions. However, they have not been adapted to photoacoustic imaging since they rely on multiple acquisitions under different controlled excitations, such as input plane-wave illuminations, which do not result in signal changes in photoacoustics. 
Here, we introduce a framework that enables the direct application of the state-of-the-art reflection-matrix based aberration correction techniques to photoacoustic imaging of dynamic targets. 
Specifically, we show that a covariance matrix analysis of  a conventional set of photoacoustic frames of dynamic targets, such as flowing red blood cells in blood vessels,  yields a virtual reflection-matrix that is mathematically analogous to a pulse-echo reflection-matrix, and lends itself to direct processing by conventional reflection-matrix based scattering-compensation algorithms. 
We validate and demonstrate the approach for photoacoustic aberration correction of vessel-mimicking targets containing flowing absorbers in both simulations and experiments. 
\end{abstract}


\maketitle 

\begin{figure*}
\centering{}\includegraphics[width=0.999\textwidth]{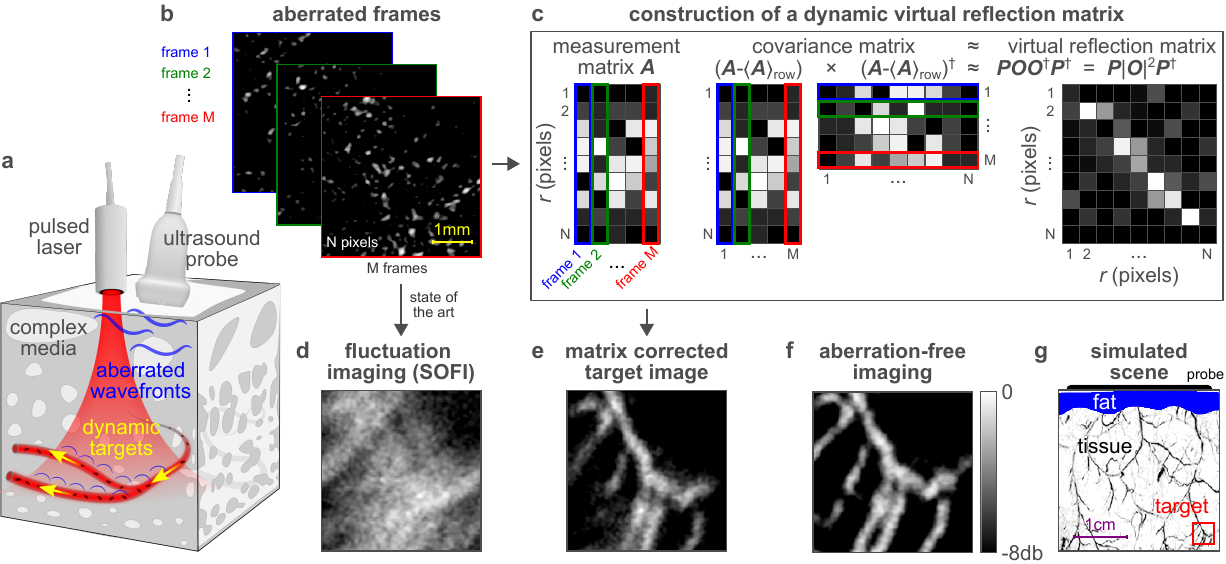}
\caption{\label{fig_concept}
Photoacoustic aberration-correction using fluctuations covariance matrix analysis -- concept illustration and numerical results. 
\textbf{a}, The experimental setup: a laser pulse (red beam) illuminates a sample containing dynamic absorbing targets (e.g., flowing red blood cells, yellow arrows), generating acoustic waves that propagate through an aberrating heterogeneous media to an ultrasound probe (blue).  
\textbf{b}, A set of photoacoustic frames of a vessel like target is acquired by sequential optical pulses. The photoacoustic frames differ from each other due to the variations in the locations of the flowing absorbers. However, all frames are distorted by the same acoustic aberrations, leading to a deterioration in resolution. The full simulated scene is illustrated in panel g, and described in Appendix~\ref{subsec_simulating_photoacoustic_data}.
\textbf{c}, Our approach: A virtual reflection matrix is constructed by reshaping the individual frames into columns of a 'measurement matrix' (left) and computing its covariance matrix (right). The virtual reflection-matrix is processed by a conventional reflection-matrix-based scattering-compensation algorithm to yield an aberration-corrected image (e).
\textbf{d}, Conventional photoacoustic fluctuation imaging: A distorted reconstruction can be obtained by plotting the temporal variance of each pixel fluctuations.
\textbf{e}, Aberration-corrected image obtained by applying CLASS reflection-matrix processing algorithm, \cite{kang_class_2017,lee_class_2022,weinberg_sunray_class_2023} on the virtual reflection matrix.
\textbf{f}, Fluctuation imaging of the same target, but in a homogeneous medium, offering a baseline for 'ground-truth' comparison. The matrix corrected target image through the aberrating layer of fat (e) closely matches the target image in this perfect imaging scenario.
\textbf{g}, The simulated scene consists of a vessel target containing flowing absorbers, surrounded by a homogeneous medium (white), and imaged through an aberrating fat layer (blue) using a linear array probe (black). The imaged field-of-view relative to the ultrasound probe is depicted by a red rectangle.
Panels b,d–f share a colorbar shown in panel f and a scale bar shown in panel b. 
}
\end{figure*}

\textit{Introduction.} Photoacoustic imaging/tomography is one of the leading techniques for deep tissue optical imaging. It allows single-shot wide field-of-view (FoV) three-dimensional reconstruction by illuminating the medium with a short optical pulse and detecting, with an acoustic detector array, the acoustic waves that are generated in the medium as result of light absorption. \cite{LV_Wang_review_2012, manohar_photoacoustics_2016}
A single-shot reconstruction is conventionally obtained by digital beam-forming (`back-propagating') the detected acoustic waves to the imaged FoV via, e.g., a delay-and-sum algorithm, under the assumption of a homogeneous and known speed-of-sound in the medium. \cite{Perrot_Garcia_2021}
When imaging in an acoustically-homogeneous medium the conventional transverse imaging resolution is the acoustic diffraction-limit, which is roughly the product of the acoustic wavelength and the array f-number. \cite{goodman2005introduction}
However, when imaging in a non-homogenous medium a key challenge is the deterioration of imaging resolution due to acoustic aberrations.  Similar to conventional ultrasound echography, acoustic aberrations result from spatial variations in the speed of sound. These lead to unknown distortions in the recorded ultrasound signals (Fig.~\ref{fig_concept}a), and distorted reconstructed images with deteriorated resolution (Fig.~\ref{fig_concept}b). In the past two decades, there have been many efforts in undoing the effects of acoustic aberrations in ultrasound echography, \cite{lambert_reflection_2020,stahli_forward_2020,bendjador_svd_2020,weber_phase_2021,poudel_iterative_2020} seismic imaging, \cite{class_geology_giraudat_2023} and photoacoustics, \cite{Pattyn_heterogeneous_2021, wang_photoacoustics_2020, modgil_geometrical_2010, jose_sos_2012, deanben2021high, dean2022practical,slobodkin_2024} as well as in optics. \cite{shemonski_computational_2015,fienup_aberration_2003,thurman_correction_2008,tippie_high-resolution_2011,tippie_multiple-plane_2010,tippie_phase-error_2009,adie_computational_2012, kang_class_2017,lee_class_2022,badon_distortion_2020, bertolotti_and_katz_2022imaging}

In ultrasound echography and photoacoustic tomography, when the variations in the speed-of-sound (SoS) are small (below 10\%), as is largely the case in soft tissues, refraction is negligible, and several modified reconstruction techniques, based on digitally compensating for the delays in time-of-flight of the recorded ultrasound waves while neglecting refraction, can be effectively applied. \cite{lambert_reflection_2020, Pattyn_heterogeneous_2021, modgil_geometrical_2010, jose_sos_2012} 
The time-of-flight delays required for correction are found in ultrasound echography using a matricial approach \cite{lambert_reflection_2020} or other processing techniques that estimate the speed-of-sound variations in the medium from ultrasound echographic measurements. \cite{Pattyn_heterogeneous_2021, jose_sos_2012} 
In scenarios that involve significant variations in the speed-of-sound, full-wave model-based algorithms can, in principle, accommodate any effect associated with acoustic propagation in strongly mismatched tissues. However, this strategy demands precise prior knowledge or estimation of the distribution of acoustic properties, and computationally intensive calculations. \cite{dean2022practical} Alternatively, if the photoacoustic targets are localized within the usually small isoplanatic patch (or memory effect) and a reference signal from a point-like source (a guide star) is available, isoplanatic  (often termed `memory-effect') based reconstruction becomes feasible. \cite{deanben2021high, dean2022practical} 

Alternatively, when a thick acoustically aberrating layer is located near the transducer array, previous works have modeled heterogeneous media as layered structures. In these works, modified ray-based and wave-based delay-and-sum algorithms, where refraction and diffraction are taken into account by applying Snell’s law \cite{wang_medical_physics_2008,wang_photoacoustics_2020} or Fresnel-Kirchhoff based models, \cite{slobodkin_2024} were used for undoing the effects of aberrations. While such modified beamforming approaches carry a lower computational cost compared to full-wave approaches, these models also require prior knowledge of the thickness profile of the aberrating layers, obtained by an additional x-ray computed tomography \cite{wang_photoacoustics_2020} or ultrasound transmit-receive acquisition, \cite{slobodkin_2024} before beamforming the target. 
Blindly retrieving the aberration and the target profile is an ill-posed problem that requires heavy regularization and/or priors. \cite{dean2022practical, slobodkin_2024} Such double-blind reconstruction was recently demonstrated for a single thick and highly aberrating layer. \cite{slobodkin_2024} Notably, in photoacoustic imaging/tomography, all the aberration-correction techniques to date consider only static targets and static aberrations. 

In optics and ultrasound, the state of the art in computational aberration and scattering compensation are reflection-matrix based approaches, which have proved extremely effective in compensating thick multiple-scattering media. \cite{kang2023tracing,haim2025image,kang_class_2017,lee_class_2022,lambert_reflection_2020,badon_distortion_2020,Kang_CLASS_Implementation_2025} 
These techniques are based on a matricial analysis of multiple measurements of the same static target, acquired under different known illuminations or sonifications, respectively.
Such methods require projecting multiple different wavefronts to the sample and recording the reflected (or transmitted) signals. The recorded signals are then used to form a reflection matrix that is, in essence, the set of measured Green's function of the medium, holding information about both the target and the aberrating media. The challenge in aberration or scattering compensation is to separate the target information and the aberrations from the measured reflection matrix.
In the past decades, several approaches to distill the target and aberration from reflection matrices in optics, acoustics, acousto-optics and seismology have been developed and experimentally demonstrated.  \cite{kang_class_2017,lee_class_2022,lambert_reflection_2020,badon_distortion_2020, class_geology_giraudat_2023,Sunray_APL_2024}
In photoacoustic imaging, a transmission-matrix approach was demonstrated for combatting \textit{optical} aberrations, enhancing light delivery with acoustic guidance, \cite{Kong2011, chaigne2014controlling} but was useful only under the assumption that no acoustic aberrations are present. 

Although extremely powerful, reflection-matrix-based aberration/scattering compensation was not utilized to correct acoustic aberrations in photoacoustic imaging. The fundamental reason is that photoacoustic tomography is inherently a single-shot technique. Changing the input illumination (e.g. angle) does not lead to a significant change in the recorded acoustic signals, since the short-coherence illumination is multiply-scattered and diffused in the sample, resulting in a rather homogeneous instantaneous illumination of the target, which remains insensitive to changes in the input illumination. 
When long-coherence light is used, optical speckle variations may be exploited for resolution improvement, \cite{chaigne_speckle_fluctuation_imaging_2016,Gateau:14,hojman_photoacoustic_2017} but the small-amplitude speckle fluctuations can be practically measured only at high ultrasound frequencies and rather shallow depths.

Nonetheless, multiple different photoacoustic measurements of the same target can be obtained when the target itself is dynamic. \cite{chaigne_flow_fluctuation_imaging_2017}  
Previous works have demonstrated that such temporal fluctuations can be used to surpass the acoustic diffraction limit and tackle the limited-view problem by analyzing higher-order temporal moments of each image pixel, such as the pixel (auto)variance. \cite{chaigne_speckle_fluctuation_imaging_2016,chaigne_flow_fluctuation_imaging_2017,vilov_fluctuation_imaging_blood_2020,Godefroy_fluctuation_imaging_3D_2023} This approach, known as photoacoustic fluctuation imaging, is the photoacoustic counterpart of super-resolution optical fluctuation imaging (SOFI), \cite{SOFI_2009} used in fluorescent microscopy by utilizing the fluctuations from blinking emitters. 

In photoacoustics, since flowing red blood cells provide a source of naturally fluctuating high contrast absorption, photoacoustic fluctuation imaging has been successfully utilized \textit{in vivo} to obtain 3D images of the vascularization in a chicken embryo. \cite{vilov_fluctuation_imaging_blood_2020,Godefroy_fluctuation_imaging_3D_2023} Nonetheless, as in conventional photoacoustics, photoacoustic fluctuation imaging is based on a delay-and-sum reconstruction assuming a homogeneous and known speed of sound, and does not compensate for any unknown aberrations induced by acoustic propagation in heterogeneous media (Fig.~\ref{fig_concept}).

\begin{figure*}
\centering{}\includegraphics[width=0.73\textwidth]{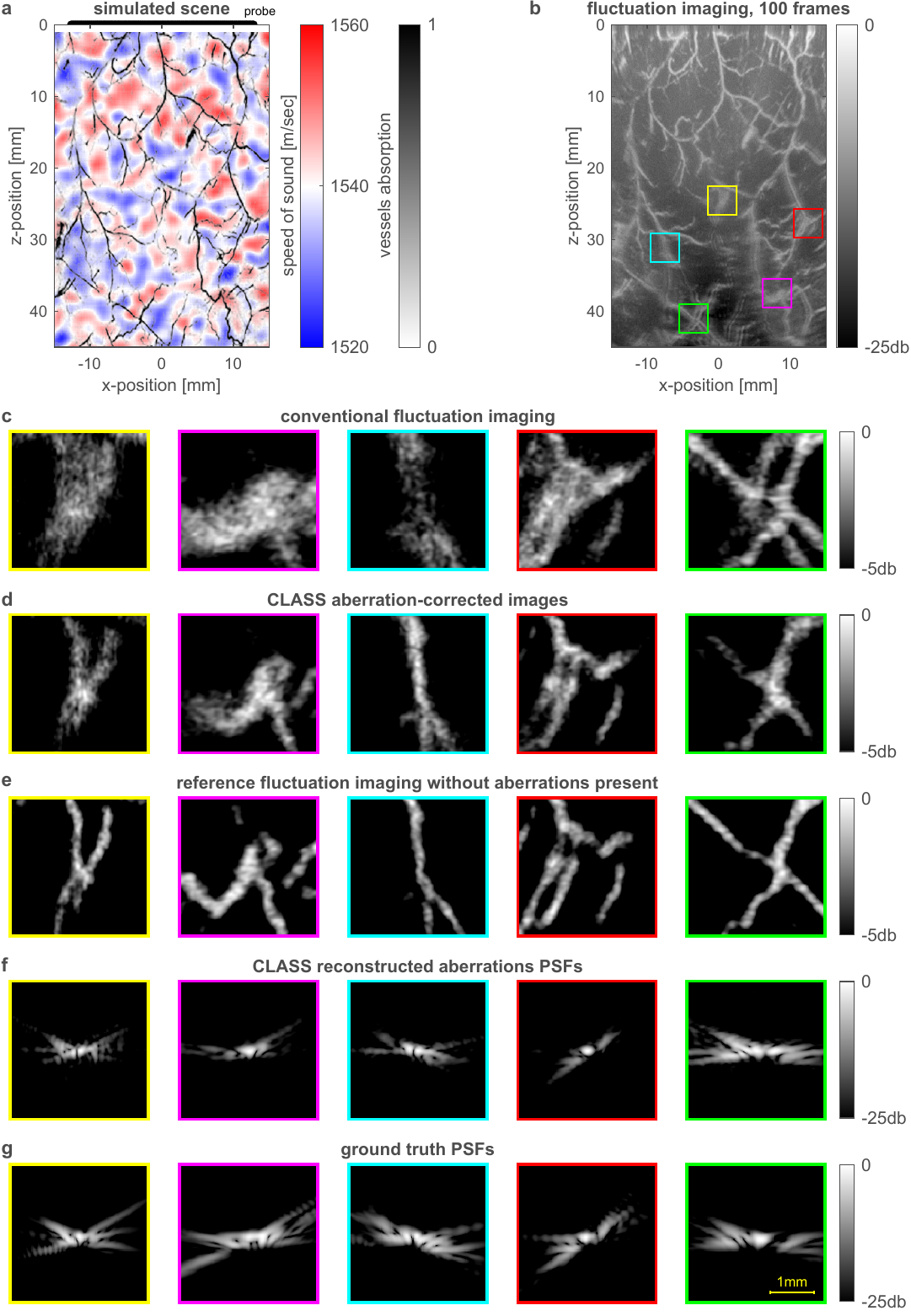}
\caption{\label{fig_simulation_unisoplanatic}
Numerical comparison between the presented photoacoustic covariance matrix approach and conventional photoacoustic fluctuation imaging in an inhomogeneous medium imposing anisoplantic aberrations.  
\textbf{a}, The simulated scene consists of a vessel-shaped target (black) containing flowing absorbers, surrounded by soft tissue and imaged using a linear array probe (black line). 
\textbf{b}, Conventional photoacoustic fluctuation imaging: the variance image of 100 frames reveals the vessel structure, but it is distorted by spatially varying aberrations (c).
\textbf{c}, Zoomed-in views of five selected regions in panel b, each corresponding to its respective color-coded window.
\textbf{d}, Reconstructions obtained in our approach: processing the photoacoustic covariance matrix of 100 frames using the CLASS algorithm \cite{weinberg_sunray_class_2023, kang_class_2017, lee_class_2022} corrects the aberrations and recovers the true vessel structure.
\textbf{e}, Conventional photoacoustic fluctuation imaging in an idealized imaging scenario without aberrations present, serving as a reference for the reconstructed images in panel d.
\textbf{f}, The spatially varying point spread function (PSFs) reconstructed by the CLASS algorithm.
\textbf{g}, k-Wave simulation of the true aberrated PSFs at the center of each selected region, providing a comparison against the reconstructed PSF in panel f.
}
\end{figure*}

\begin{figure*}
\centering{}\includegraphics[width=0.999\textwidth]{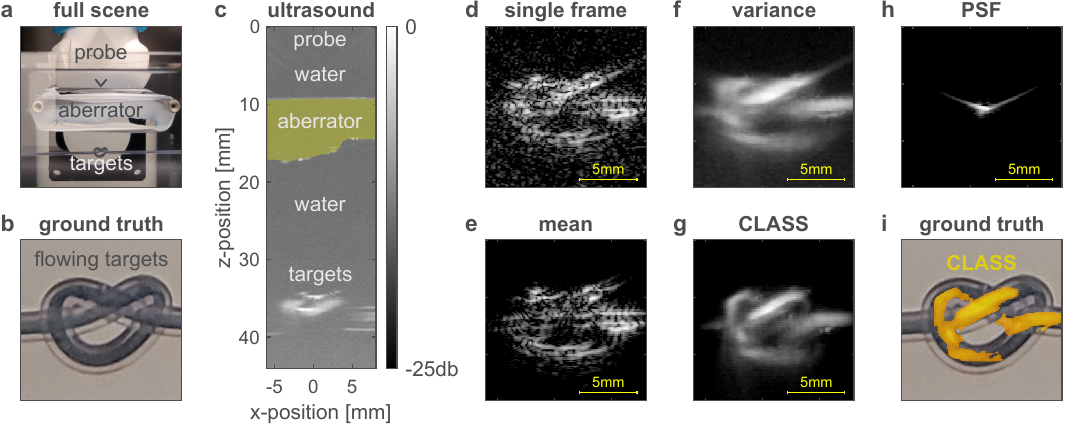}
\caption{\label{fig_experimental_knot}Experimental photoacoustic covariance-matrix imaging through a tissue-mimicking aberrator. 
\textbf{a}, Experimental setup: a linear array ultrasound probe images a vessel like structure composed of silicon tubes with flowing absorbers (6µm diameter polymer microspheres) through an agar-based tissue-mimicking aberrator (aberrator thickness ranges between 5-8mm). 
\textbf{b}, Optical image of the vessel-like tubes containing flowing absorbing dye solution. 
\textbf{c}, Ultrasound echography image shows the aberrating layer and an aberrated image of the silicon tubes.
\textbf{d}, A single photoacoustic image of the targets, displaying an aberrated shape of the tubes. Only top and bottom parts are visible due to the probe limited view.
\textbf{e}, Averaged 200 frames display higher SNR but no clearer features or resolution. 
\textbf{f}, Conventional fluctuation-imaging (auto)variance analysis provides an aberrated view of the flowing absorbers inside the tubes. 
\textbf{g}, Covariance matrix aberration-corrected image, reconstructed by applying the CLASS algorithm \cite{weinberg_sunray_class_2023, kang_class_2017, lee_class_2022} to the photoacoustic covariance matrix, showing an improved reconstruction of the flowing absorbers distribution in the true inner cross-section of the tubes.
\textbf{h}, The aberrated point spread function, retrieved by the aberration-correction CLASS algorithm.
\textbf{i}, Overlay of the aberration-corrected image from panel g (yellow) on the true shape of the tubes.
Panels c-h share a colorbar shown in panel c.}
\end{figure*}

Here, we introduce a novel computational framework that enables the direct application of the state-of-the-art reflection-matrix-based aberration correction techniques \cite{lee_class_2022} to photoacoustic imaging of dynamic optical absorbers, such as flowing red blood cells in blood vessels, with a conventional photoacoustic imaging setup, and without the use of any additional echographic or other non-photoacoustic measurements. 
Specifically, we show that the cross-correlations of temporal fluctuations between pixels in a set of aberrated photoacoustic frames, form a covariance matrix that has the same mathematical form as a `virtual' reflection matrix (Fig.~\ref{fig_concept}c). The mathematical equivalence between the photoacoustic covariance matrix and the ultrasound reflection matrix results in the covariance matrix lending itself to direct processing by conventional reflection-matrix based algorithms. 
We demonstrate these analytic predictions both numerically and experimentally by performing reflection-matrix based computational photoacoustic aberration correction of vessel-mimicking targets containing flowing absorbers. We show that analyzing the photoacoustic covariance matrix yields both the corrected photoacoustic image and the aberrated point spread function (PSF) (Fig.~\ref{fig_concept}-\ref{fig_experimental_knot}). 

\textit{Principle.} Our approach is performed in three steps  (Fig.~\ref{fig_concept}): the first is a conventional photoacoustic acquisition of multiple frames of a dynamic fluctuating absorbing medium, such as flowing blood. The dataset, which is the same dataset used in photoacoustic fluctuation imaging \cite{chaigne_flow_fluctuation_imaging_2017} is then used to construct a photoacoustic covariance matrix that is then processed by conventional matricial scattering compensation algorithms. Here we apply the `closed-loop accumulation of
single scattering' (CLASS) algorithm. \cite{lee_class_2022} 

\textit{Constructing the covariance virtual reflection matrix}. Following signal acquisition and beamforming (Supplemental Material, Appendix~\ref{subsec_experimental_work}) of a stack of aberrated photoacoustic frames (Fig.~\ref{fig_concept}b), a virtual reflection matrix is formed from the cross-correlations between the different image pixels (Fig.~\ref{fig_concept}c). This is performed by first reshaping the beamformed images into columns of a `measurement matrix’, $\boldsymbol{\it{A}}$, with each column representing a single complex-valued photoacoustic frame. The covariance matrix, $\boldsymbol{\it{R}}$, is defined as the cross-correlation of the measurement matrix after subtraction of its row-wise (temporal) mean: 
\begin{equation}\label{eq_covariance_definition}
    \boldsymbol{\it{R}} \equiv \left(\boldsymbol{\it{A}}-\left<\boldsymbol{\it{A}}\right>_\text{row}\right) \left(\boldsymbol{\it{A}}-\left<\boldsymbol{\it{A}}\right>_\text{row}\right) ^ \dagger \equiv \hat{\boldsymbol{\it{A}}} \hat{\boldsymbol{\it{A}}} ^ \dagger
\end{equation}
Following the recent works in optics, \cite{lee_class_2022, weinberg_sunray_class_2023} this covariance matrix can be interpreted as a `virtual reflection matrix', where each pixel effectively serves as both a detector and a virtual acoustic point source, as we show below. This result bears an analogy to Green function retrieval in passive correlation imaging. \cite{derode2003recovering,wapenaar2004retrieving,garnier2016passive} 
Interestingly, the diagonal of the covariance matrix is the conventional temporal-variance of each pixel, previously used to improve resolution and to address the limited-view artifacts in photoacoustic fluctuation imaging. \cite{chaigne_flow_fluctuation_imaging_2017,vilov_fluctuation_imaging_blood_2020,Godefroy_fluctuation_imaging_3D_2023} 
The off-diagonal, cross-correlations terms, which are naturally available in any photoacoustic image stack of a dynamic sample, were not utilized to date in photoacoustic imaging. Our results below demonstrate how  acoustic aberration correction can be performed using these cross-correlations terms.
We note that while we chose here to calculate the covariance matrix in the reconstructed image pixel domain (or `basis'), a similar calculation can be performed on the raw spatiotemporal RF signals, resulting in a covariance matrix in the spatiotemporal basis.
The covariance matrix equivalence to a virtual reflection matrix has been discussed in detail elsewhere, \cite{weinberg_sunray_class_2023} where the covariance matrix is similarly used as a virtual reflection matrix for incoherent optical imaging. A similar mathematical formulation and analogy can be drawn for photoacoustic imaging: This formulation begins by expressing the measurement matrix,  $\boldsymbol{\it{A}}$,  as a multiplication of three (initially unknown) matrices: the dynamic absorbers distribution $\boldsymbol{\it{S}}$, the target fixed support, $\boldsymbol{\it{O}}$, and the aberration PSF, $\boldsymbol{\it{P}}$:
\begin{equation}\label{eq_measurement_matrix_definition}
    \boldsymbol{\it{A}} = \boldsymbol{\it{P}}\boldsymbol{\it{O}}\boldsymbol{\it{S}}
\end{equation}
where $\boldsymbol{\it{S}}$ is a random matrix with each of its columns containing the random (unknown) positions of the flowing optical absorbers (e.g., red blood cells) at the target plane, $\boldsymbol{\it{O}}$ is a diagonal matrix with the fixed support of the target absorption (i.e., the blood vessels structure) as its diagonal elements, \cite{hojman_photoacoustic_2017} and $\boldsymbol{\it{P}}$ is a convolution matrix representing the unknown distorted acoustic detection PSF, i.e., a matrix with columns containing the acoustic PSF for the different points in the imaged field of view. 

Notably, using the decomposition of Eq.~\eqref{eq_measurement_matrix_definition}, it becomes apparent that subtracting the row mean from the known measurement matrix $\boldsymbol{\it{A}}$ –- that is, removing the temporal mean value from each target pixel across $M$ realizations/frames of the flowing optical absorbers –- is equivalent to subtracting the row mean from the random (unknown) dynamic absorbers distribution:
\cite{weinberg_sunray_class_2023}
\begin{equation}\label{eq_measurement_matrix_hat_definition}
    \hat{\boldsymbol{\it{A}}} = \boldsymbol{\it{A}} - \langle \boldsymbol{\it{A}}\rangle_\text{row} = \boldsymbol{\it{P}} \boldsymbol{\it{O}} \left(\boldsymbol{\it{S}} - \langle\boldsymbol{\it{S}}\rangle_\text{row} \right)
    \equiv \boldsymbol{\it{P}} \boldsymbol{\it{O}} \hat{\boldsymbol{\it{S}}}
\end{equation}

Plugging Eq.~\eqref{eq_measurement_matrix_hat_definition} into Eq.~\eqref{eq_covariance_definition}, the covariance matrix $\boldsymbol{\it{R}}$ takes the form (Supplemental Material, Appendix~\ref{sec_CLASS_algorithm}):
\begin{equation}\label{eq_covariance_as_virtual_reflection}
    \boldsymbol{\it{R}} = \boldsymbol{\it{P}} \boldsymbol{\it{O}} \hat{\boldsymbol{\it{S}}} \hat{\boldsymbol{\it{S}}}^\dagger \boldsymbol{\it{O}}^\dagger \boldsymbol{\it{P}}^\dagger \approx \boldsymbol{\it{P}} \boldsymbol{\it{O}} \boldsymbol{\it{O}}^\dagger \boldsymbol{\it{P}}^\dagger = \boldsymbol{\it{P}} \left|\boldsymbol{\it{O}}\right|^2 \boldsymbol{\it{P}}^\dagger
\end{equation}
where we have used the fact that the fluctuations between different pixels are uncorrelated, i.e.: $\hat{\boldsymbol{\it{S}}} \hat{\boldsymbol{\it{S}}}^\dagger\approx \boldsymbol{\it{I}}$. 
This form is equivalent to a conventional acoustic reflection matrix in the case of acoustic aberrations described by $\boldsymbol{\it{P}}$. \cite{lambert_reflection_2020}  The goal of an aberration-compensation noninvasive imaging technique is to retrieve $\boldsymbol{\it{O}}$ (and potentially also $\boldsymbol{\it{P}}$) from $\boldsymbol{\it{R}}$, where $\boldsymbol{\it{P}}$, $\boldsymbol{\it{O}}$ and $\boldsymbol{\it{S}}$ are unknown. This is exactly what the current state-of-the-art reflection-matrix based scattering-compensation algorithms  \cite{lambert_reflection_2020,lee_class_2022,weinberg_sunray_class_2023} are designed for, as we explain below.

\textit{Reflection-matrix-based aberration correction}. The virtual reflection matrix, $\boldsymbol{\it{R}}$, can be analyzed and processed by any of the well-established aberration-correction techniques developed for the conventional reflection matrix in optics or ultrasound. \cite{kang_class_2017,lee_class_2022,badon_distortion_2020,lambert_reflection_2020} As a demonstration, we apply the memory-efficient CLASS iterative phase-correction algorithm, \cite{weinberg_sunray_class_2023} to effectively correct the effects of isoplanatic acoustic aberrations from the photoacoustic dataset (Fig.~\ref{fig_concept}-\ref{fig_experimental_knot}). We note that this memory-efficient implementation directly calculates the aberration correction from the measured dataset without the need to explicitly calculate the full covariance matrix, which has a very large number of elements (the number of image pixels squared). A brief  intuitive description of the CLASS aberration estimation algorithm is presented in Appendix~\ref{sec_CLASS_algorithm}.

\textit{Numerical results.} To validate the concept of photoacoustic aberration correction using the covariance matrix of natural flucutations, we performed both numerical and experimental studies. Fig.~\ref{fig_concept} and Fig.~\ref{fig_simulation_unisoplanatic} present the results of numerical simulations of photoacoustic imaging of human vasculature in complex tissue-mimicking media. Fig.~\ref{fig_concept} shows isoplanatic  aberration correction for a simulated target of small vessels imaged through an aberrating layer of fat (panel g). Fig.~\ref{fig_simulation_unisoplanatic} focuses on the reconstruction of simulated multiple vessels targets with a realistic soft tissue-mimicking speed of sound distribution (panel a).

In each case, a series of 100 photoacoustic frames was generated, each corresponding to a different random distribution of absorbing red blood cells in the vessel target. 

The introduction of the simulated fat aberrator (Fig.~\ref{fig_concept}g) led to considerable distortions in the photoacoustic images (Fig.~\ref{fig_concept}b,d). Computing the variance of the frame sequence, known as photoacoustic fluctuation imaging, \cite{chaigne_speckle_fluctuation_imaging_2016,chaigne_flow_fluctuation_imaging_2017,vilov_fluctuation_imaging_blood_2020,Godefroy_fluctuation_imaging_3D_2023} fails to reconstruct the branching vessel structure (Fig.~\ref{fig_concept}d).

As a demonstration for matrix-based computational aberration-correction, we applied the memory-efficient closed-loop accumulation of single scattering (CLASS) algorithm \cite{weinberg_sunray_class_2023} to the photoacoustic covariance matrix to correct for isoplanatic aberrations. The aberration-corrected image (Fig.~\ref{fig_concept}e) demonstrated successful recovery of the vessel structure and compensation of the aberration, closely matching the variance image obtained without the aberrator present (Fig.~\ref{fig_concept}f).

In Fig.~\ref{fig_simulation_unisoplanatic}, we extend our numerical study to a more realistic imaging scenario, where multiple vessel-shaped targets are embedded within a tissue-mimicking medium with a spatially varying speed of sound (Fig.~\ref{fig_simulation_unisoplanatic}a). Conventional photoacoustic fluctuation imaging, based on computing the variance of 100 frames, reveals the vessel structures but suffers from significant image distortions due to spatially varying aberrations (Fig.~\ref{fig_simulation_unisoplanatic}b). The effect of these aberrations is further illustrated in the zoomed-in views of five selected regions (Fig.~\ref{fig_simulation_unisoplanatic}c), where the vessel shapes appear smeared, duplicated, and shifted.

To overcome these aberrations, we applied the CLASS algorithm \cite{weinberg_sunray_class_2023, kang_class_2017, lee_class_2022} to the photoacoustic covariance matrix. The resulting aberration-corrected image (Fig.~\ref{fig_simulation_unisoplanatic}d) successfully restores the vessel structures, closely matching the reference images obtained in an idealized scenario without aberrations (Fig.~\ref{fig_simulation_unisoplanatic}e). Furthermore, to analyze the nature of the aberrations, we reconstructed the spatially varying point spread function (PSF) using CLASS (Fig.~\ref{fig_simulation_unisoplanatic}f) and compared it to \textit{k-Wave} simulations \cite{k_wave_toolbox_2010} of the PSF at the center of each selected region, obtained by imaging a point absorber (Fig.~\ref{fig_simulation_unisoplanatic}g). The agreement between the reconstructed and simulated PSFs further supports the accuracy of the aberration correction process.

\textit{Experimental results.} Beyond numerical validation, we demonstrated the approach experimentally using a tissue-mimicking phantom and a conventional linear-array photoacoustic system. 
Experimental results are shown in Fig.~\ref{fig_experimental_knot}, and additional results are available in Appendix~\ref{sec_experimental_additonal_results}. 
The tubes containing the flowing optical absorbers are visible in Fig.~\ref{fig_experimental_knot}a-b.
A sequence of photoacoustic frames was acquired using an 8MHz linear array transducer, and illumination by 6ns pulses at 670nm, while optically-absorbing beads were flowing through the tubes. 

A single frame (Fig.~\ref{fig_experimental_knot}d) showed mainly the top and bottom boundaries of the tubes, which added an additional static photoacoustic signal to all frames. As a result of the limited-view of the probe, the average of all frames showed only the top and bottom boundaries of each tube  (Fig.~\ref{fig_experimental_knot}e). On average, the signals originating from the fluctuating optical absorbers as well as the side boundaries of the tubes destructively interfere, leading to no signal inside the tubes. 

As expected, a conventional photoacoustic fluctuation imaging variance image provides an aberrated view of the absorbers distribution inside the tubes, with distortions and artifacts caused by the unaccounted for aberrations of the tissue-mimicking layer and the walls of the tubes (Fig.~\ref{fig_experimental_knot}f).

The photoacoustic covariance matrix was constructed from the sequence of aberrated single frames and processed using the CLASS algorithm. The resulting aberration-corrected  image (Fig.~\ref{fig_experimental_knot}g) provides an estimation of the true distribution of the flowing absorbers inside the tubes, visible by the correct cross-section of the tubes inner volume (Fig.~\ref{fig_experimental_knot}i). CLASS also provides the aberrated PSF (Fig.~\ref{fig_experimental_knot}h).

\textit{Discussion.} In conclusion, we have presented a novel approach for computationally correcting acoustic aberrations in photoacoustic imaging of dynamic targets. 
Our approach exploits natural fluctuations, such as blood flow, to construct a reflection-matrix-like data from a conventional photoacoustic imaging system. The only requirement for the approach is that spatially uncorrelated temporal fluctuations exist. This can be achieved either by true physical fluctuations, as demonstrated here, or alternatively by illumination induced fluctuations. Importantly, no additional information or assumptions or any change in conventional photoacoustic imaging hardware is required. The approach extends photoacoustic fluctuations imaging to aberration correction by taking into account the rich information contained in the off-diagonal terms of the photoacoustic covariance matrix, i.e., in the cross-correlations of the fluctuations between different imaged pixels, in a similar fashion to passive cross-correlation imaging. \cite{garnier2016passive}

We have considered the case of dynamic flow under static aberrations. In this case, once the aberration PSF is found, aberration-corrected photoacoustic imaging can be performed in real time by applying the known correction. Importantly, the aberration correction can be applied on each single frame to visualize the flow dynamics. When the aberration itself is dynamic, the dataset may be divided to several datasets of relatively constant aberrations, with each dataset processed separately to retrieve the corrected images.

To image a large FoV where the aberrated PSF varies across the imaged field of view, we performed anisoplanatic aberration correction by dividing the FoV and separately correcting smaller FoVs that can be mosaicked to a single large image. \cite{alterman2021imaging,lee_class_2022,najar23} Alternatively, recent works in reflection-matrix based imaging address the issue of spatially varying PSFs by considering a model of a layered medium in the matricial processing. \cite{kang2023tracing, lambert20} A similar very effective model was recently demonstrated by an image-quality optimization based approach. \cite{haim2025image} This approach can be adapted to enhance the proposed covariance matrix imaging framework.

Another area for further development is the integration of additional reflection-matrix-based techniques from optics and ultrasound such as image scanning microscopy (ISM) or pixel reassignment, \cite{ISM_2010,Sheppard_pixel_reassignment_2020,Sommer_pixel_reassignment_2021} into the photoacoustic covariance-matrix framework, to improve resolution and SNR. 

Ultimately, applying photoacoustic covariance-matrix imaging to \textit{in vivo} imaging applications, where the motion of blood cells provides a natural source of fluctuations, would require handling fluctuations from other sources such as breathing, movements, and possibly dynamic aberrations, as is performed in conventional photoacoustic fluctuation imaging. \cite{Godefroy_fluctuation_imaging_3D_2023, vilov_fluctuation_imaging_blood_2020} 

\textit{Acknowledgement.} This project has received funding from the European Research Council under the European Union’s Horizon 2020 Research and Innovation Program grant number 101002406.
 
\textit{Data availability}. The data that support the findings of this study are available from the corresponding author upon reasonable request.

\section*{Supplementary Material:\\
Leveraging Natural Fluctuations for Matrix-Based Aberration Correction in Photoacoustic Imaging} 

\appendix

\section{Numerical simulation}\label{subsec_simulating_photoacoustic_data}

The simulated scenario is presented in Fig.~\ref{fig_concept}g and Fig.~\ref{fig_simulation_unisoplanatic}a. Both simulations represent photoacoustic imaging of human vasculature in complex tissue-mimicking media. 
The solid line at the top of each panel (Fig.~\ref{fig_concept}g, Fig.~\ref{fig_simulation_unisoplanatic}a) marks the position of a 128-detector linear array (probe). A vessels-like target is located $\sim$2.5-4cm away from the probe and surrounded by soft tissue. Initial pressure distribution is generated by multiplying the vessels structure with a random matrix, representing random positions of flowing optical absorbers (red blood cells). This initial pressure distribution is generated 100 times with different random positions of the flowing red blood cells. 

For the numerical study, the RF-data for the set of 100 photoacoustic frames was simulated using the \textit{k-wave} toolbox. \cite{k_wave_toolbox_2010} For each simulated frame, an initial pressure field from a random distribution of 10µm-diameter absorbers with 50\% volume filling fraction was propagated through a heterogeneous medium by iteratively solving the continuity equation (conservation of mass) and Euler's equation (conservation of momentum), assuming a linear adiabatic equation of state. 
Fig.~\ref{fig_concept} simulates a scenario where a 4mm-thick layer of fat near the probe refracts and delays the ultrasound waves as they propagate from the vessels target to the probe. 
For this simulation, soft tissue was simulated as homogeneous media with speed of sound 1540m/s, as in diagnostic ultrasound imaging. \cite{speed_of_sound_in_soft_tissue} The aberrating layer was simulated using the acoustic properties of fat \cite{tissues_database_2022} (speed of sound 1440m/s and mass density 911kg/m\textsuperscript{3}). In Fig.~\ref{fig_simulation_unisoplanatic}, soft tissue mimicking speed of sound distribution was generated by a squared exponential distribution as described elsewhere. \cite{Tick_2020} 
A cropped version of a photoacoustic image of human calf vasculature, adapted from Asao et al., \cite{PA_human_calf2022} served as the vessel-mimicking target in all simulations, under the terms of the CC BY-NC 4.0 license.  
The RF-data at the locations of the sensors (probe in Fig.~\ref{fig_concept}g and Fig.~\ref{fig_simulation_unisoplanatic}a) was filtered with a gaussian filter, centered at 15MHz with 80\% bandwidth, typical values for medical ultrasound transducers used in our experiments. A Gaussian random white noise with mean-to-mean signal-to-noise ratio (SNR) of 10.0db, corresponding to noise intensity equal to 10\% signal intensity, was added to all signals before applying band-pass filtering and reconstruction algorithms.
To mimic experimental conditions, a different random time delay was introduced to all elements signals in each recording to simulate realistic jitter between multiple photoacoustic measurements of the same target. The simulated jitter had a standard deviation of $40ns$, a value similar to the jitter in our experimental system (see below). A detailed description of computation times, memory requirements and grid sizes appears in Appendix~\ref{sec_implementation}.

\section{Experimental data acquisition}\label{subsec_experimental_work}

To experimentally demonstrate our approach, we performed photoacoustic experiments in a water tank setup using a conventional photoacoustic imaging system. The optical source was a pulsed laser (InnoLas Inno P1864, 6ns pulse duration, 100Hz repetition rate, 670nm wavelength, 25mJ per pulse). 
A 128-detector linear array (Verasonics L12-3V transducer) connected to a 256 channels high frequency research ultrasound system (Verasonics Vantage 256) was used for ultrasound echography and photoacoustic acquisitions.

Vessel-mimicking targets were fabricated by flowing water with  optically absorbing 6µm diameter polymer microspheres (Polybead, catalog no. 24293-5), imitating red blood cells, through transparent silicone tubes (Quickun Pure Silicone Tubing), having 1mm inner diameter and 2mm outer diameter. 
A nearly constant rate flow of 0.5ml/min was realized using a syringe pump (KD Scientific LEGATO 110, catalog no. 788110). 
A vessel-like target was created by tying the tubes into a knot (Fig.~\ref{fig_experimental_knot}a,b) and imaging the target through a 5-8mm thick agar-based layer (Fig.~\ref{fig_experimental_knot}a,c) to introduce acoustic aberrations. This aberrating layer was made of a non-commercial agar-based tissue-mimicking material according to the preparation protocol described elsewhere. \cite{agar_prep_2001}
In an additional experiment, depicted in Fig.~\ref{fig_experimental_3_circles}, three tubes were arranged such that they crossed nearly-perpendicularly through the imaging plane at depths of 31-37mm, and imaged through a 4-5mm thick agar-based layer.

\section{Reflection-matrix-based aberration correction}\label{sec_CLASS_algorithm}

The reflection-matrix formalism provides a unified framework to describe wave propagation, object interaction, and system aberrations. In this section, we outline the mathematical model of the conventional reflection matrix, show its equivalence to the covariance-based virtual reflection matrix, and then present an intuitive overview of the CLASS algorithm \cite{kang_class_2017,lee_class_2022,weinberg_sunray_class_2023,Sunray_APL_2024,Kang_CLASS_Implementation_2025} used for aberration correction.

\subsection{Continuous-space model}
We denote by $E^{in}_m(r)$ the incident field in the $m$-th illumination, and by $E^{out}_m(r)$ the corresponding measured field at position $r$. The object scattering or absorption distribution is described by $O(r)$, while $P^{ill}(r)$ and $P^{det}(r)$ are the field point spread functions (PSFs) for illumination and detection, respectively, describing the propagation from the input plane to the object plane and vice versa, respectively. The forward model under the isoplanatic assumption gives:
\begin{equation}\label{eq:E_out_real_space}
E_{m}^{out}(r) = P^{det}(r) \circledast \bigl[O(r) \cdot E^{ill}_m(r)\bigr],
\end{equation}
where $E^{ill}_m(r)$ is the illumination field at the object plane, given by:
\begin{equation}\label{eq:E_ill_real_space}
E^{ill}_m(r) = P^{ill}(r) \circledast E^{in}_m(r).
\end{equation}

Plugging Eq.~\eqref{eq:E_ill_real_space} into Eq.~\eqref{eq:E_out_real_space}:
\begin{equation}\label{eq:E_out_E_in_relations_in_real_space}
E_{m}^{out}(r) = P^{det}(r) \circledast \Bigl[O(r) \cdot \bigl(P^{ill}(r) \circledast E^{in}_m(r)\bigr)\Bigr].
\end{equation}

Fourier-transforming to the spatial-frequency domain (denoted by tilde), convolutions become products and vice versa. We obtain:
\begin{equation}\label{eq:E_out_E_in_relations_in_k_space}
\tilde{E}^{out}_m(k) = \tilde{P}^{det}(k) \bigl[\tilde{O}(k) \circledast \bigl(\tilde{P}^{ill}(k) \cdot \tilde{E}^{in}_m(k)\bigr)\bigr].
\end{equation}

Eq.~\eqref{eq:E_out_E_in_relations_in_real_space} and Eq.~\eqref{eq:E_out_E_in_relations_in_k_space} represent the fundamental relations between any input field and measured output field in an isoplanatic linear imaging scenario. This linear relationship is often described by a set of Green functions, which can be discretized by spatial sampling to form the medium's reflection matrix, as is explained below.

\subsection{The reflection matrix formalism}

Since Eq.~\eqref{eq:E_out_E_in_relations_in_k_space} describes a linear relationship between the output and input field, the output field for any input field is the linear superposition of the output fields for the plane waves (or any other basis modes) that represent the input field. Thus, it is sufficient to study the relationship of input and output fields for plane-wave inputs: 
\begin{equation}\label{eq:plane_wave_basis_def}
\tilde{E}^{in}_m(k)=\int \tilde{E}^{in}_m(k_{in}) \delta(k - k_{in}) \;dk_{in}.
\end{equation}

Plugging Eq.~\eqref{eq:plane_wave_basis_def} to Eq.~\eqref{eq:E_out_E_in_relations_in_k_space} yields the output field as a superposition of plane wave input fields:
\begin{equation}\label{eq:E_out_E_in_relations_in_plane_wave_basis}
\tilde{E}^{out}_m(k) = \int \bigl[\tilde{P}^{det}(k) \tilde{O}(k-k_{in}) \tilde{P}^{ill}(k_{in}) \bigr] \tilde{E}^{in}_m(k_{in}) \;dk_{in}.
\end{equation}
We can therefore describe the response/transfer function as a function of input and output spatial frequencies:
\begin{equation}\label{eq:R_integral_form}
\tilde{R}(k, k_{in}) = \tilde{P}^{det}(k) \tilde{O}(k - k_{in}) \tilde{P}^{ill}(k_{in}).
\end{equation}

Eq.~\eqref{eq:E_out_E_in_relations_in_plane_wave_basis} and Eq.~\eqref{eq:R_integral_form} can be understood as a generalization of the diffraction-grating equation in free space: $k_{out}=k_{in}+k_{grating}$, where for every input plane wave at $k_{in}$ the resulting output field is the diffraction of the input field by the object (generalized grating) angular-spectrum component at $\Delta k = k-k_{in}$. In addition, propagation, aberration and scattering, to the object and from the object, that are represented by the illumination and detection PSFs, respectively, are manifested by multiplication of the input field by the input aberration at $k_{in}$ and output aberration at $k$.

Discretizing both $k$ and $k_{in}$ into indices $m$ and $n$ leads to the reflection matrix, with each of its elements $\tilde{R}_{m,n}$ given by:
\begin{equation}\label{eq:R_matrix_elements}
\tilde{R}_{m,n} = \tilde{P}^{det}_m \tilde{O}_{m,n} \tilde{P}^{ill}_n,
\end{equation}
this can be written in matricial form as:
\begin{equation}\label{eq:reflection_matrix_k_space}
    \tilde{\boldsymbol{\it{R}}} = \tilde{\boldsymbol{\it{P}}}^{det} \tilde{\boldsymbol{\it{O}}} \tilde{\boldsymbol{\it{P}}}^{ill\dagger},
\end{equation}
where $\tilde{\boldsymbol{\it{P}}}^{det}$ and $\tilde{\boldsymbol{\it{P}}}^{ill}$ are diagonal matrices (system transfer functions) in $k$-space, and $\tilde{\boldsymbol{\it{O}}}$ is a convolutional (Toeplitz) matrix capturing the object's spatial coupling in Fourier space.

Inverse Fourier transforming back to spatial coordinates recovers:
\begin{equation}\label{eq:reflection_matrix_real_space}
    \boldsymbol{\it{R}} = \boldsymbol{\it{P}}^{det} \boldsymbol{\it{O}} \boldsymbol{\it{P}}^{ill\dagger},
\end{equation}
where now $\boldsymbol{\it{O}}$ is a diagonal matrix in the spatial coordinates basis (at the object plane) and $\boldsymbol{\it{P}}^{det}, \boldsymbol{\it{P}}^{ill}$ are convolution (Toeplitz) matrices, where each column represents the shifted illumination and detection distorted PSFs.

The matrix $\boldsymbol{\it{R}}$ in Eq.~\eqref{eq:reflection_matrix_real_space} has the same Toeplitz-diagonal-Toeplitz structure as the photoacoustic `virtual reflection matrix' described in Eq.~\eqref{eq_covariance_as_virtual_reflection} of the main text, obtained by calculating the covariance matrix of a set of photoacoustic frames. This equivalence between the conventional reflection matrix (Eq.~\eqref{eq:reflection_matrix_real_space}) and the covariance-based virtual reflection matrix (Eq.~\eqref{eq_covariance_as_virtual_reflection}) provides the theoretical foundation for applying established matrix-based aberration correction algorithms such as CLASS, \cite{kang_class_2017} as we have demonstrated in our work. 
As we explain in the next section, CLASS estimates and corrects the phase distortions in the Fourier domain by an iterative procedure that leverages the diagonal-Toeplitz-diagonal structure of the reflection matrix in the Fourier domain.

\subsection{Aberration estimation and correction via CLASS}

CLASS \cite{kang_class_2017,lee_class_2022,weinberg_sunray_class_2023,Sunray_APL_2024,Kang_CLASS_Implementation_2025} is an iterative algorithm that operates on any input matrix having the form described in Eq.~\eqref{eq:reflection_matrix_k_space}, $\tilde{\boldsymbol{\it{R}}} = \tilde{\boldsymbol{\it{P}}}^{det} \tilde{\boldsymbol{\it{O}}} \tilde{\boldsymbol{\it{P}}}^{ill\dagger}$, where $\tilde{\boldsymbol{\it{O}}}$ is a Toeplitz matrix that is multiplied from both sides by diagonal matrices $\tilde{\boldsymbol{\it{P}}}^{det}$ and $\tilde{\boldsymbol{\it{P}}}^{ill}$, which represent the detection and illumination PSFs, respectively. In the case of the covariance matrix (virtual reflection matrix) considered in this work  $\tilde{\boldsymbol{\it{P}}}^{ill}=\tilde{\boldsymbol{\it{P}}}^{det}$.
The CLASS algorithm takes advantage of this unique structure of the matrix $\tilde{\boldsymbol{\it{R}}}$, and estimates $\tilde{\boldsymbol{\it{P}}}^{det}$  and $\tilde{\boldsymbol{\it{O}}}$  by calculating the correlations between the columns of $\tilde{\boldsymbol{\it{R}}}$ after appropriate relative shifting as we explain below. 

The reason why correlations between columns (or rows) of $\tilde{\boldsymbol{\it{R}}}$ can provide an estimate of the illumination aberrations ($\tilde{\boldsymbol{\it{P}}}^{ill}$) or detection aberrations, is that since $\tilde{\boldsymbol{\it{P}}}^{ill}$ is a diagonal matrix, multiplying $\tilde{\boldsymbol{\it{O}}}$ by it from the right results in that $\tilde{\boldsymbol{\it{O}}} \tilde{\boldsymbol{\it{P}}}^{ill}$ has the same Toepelitz structure of $\tilde{\boldsymbol{\it{O}}}$ with only each column, $n$, being multiplied by a single illumination (input) aberration factor $\tilde{\it{P}}^{ill}_{nn}$ (the multiplication by $\tilde{\boldsymbol{\it{P}}}^{det}$ from the left side of $\tilde{\boldsymbol{\it{O}}}$ has the same effect on the rows of $\tilde{\boldsymbol{\it{R}}}$). Thus, estimating the difference in input aberrations between two neighboring columns can be done, assuming that the detection (output) aberrations have some finite correlation range, by comparing the relative amplitudes and phases of these two neighboring columns, after shifting one of them by a single row, i.e. by correlating two shifted columns of $\tilde{\boldsymbol{\it{R}}}$. This is the basic process of a single iteration in CLASS. The entire CLASS algorithm can be described as follows:

\begin{enumerate}
    \item In each iteration, the algorithm first shifts each column of the matrix $\tilde{\boldsymbol{\it{R}}}$ to compensate for the Toeplitz (convolution) structure:  the $n$-th column is shifted by $n$ elements. After this process, in the case that no aberrations are present, all of the columns of $\tilde{\boldsymbol{\it{R}}}$ are identical. In the case that only illumination aberrations are present, all of the columns of $\tilde{\boldsymbol{\it{R}}}$ are identical except for a different single scalar multiplication for each column, which is the input aberration for this column. In this case (i.e. neglecting detection aberrations) one can thus estimate the relative illumination aberrations from cross-correlating the different columns.
    \item  Illumination aberration estimation: the illumination aberration (the values of the elements of the diagonal of $\tilde{\boldsymbol{\it{P}}}^{ill}$) is directly estimated from the correlations between each column and a neighboring column, or alternatively a fixed column or the average of the columns of $\tilde{\boldsymbol{\it{R}}}$. \cite{Kang_CLASS_Implementation_2025}
    \item Applying the illumination-aberration correction: A conjugate phase $\tilde{\boldsymbol{\it{P}}}^{ill*}$ is applied to each column to cancel the aberrations in illumination.
    \item Estimation and compensation for aberrations in detection: The matrix $\tilde{\boldsymbol{\it{R}}}$ is transpose-conjugated and the same procedure described in steps 1-3 is repeated to estimate the values along the diagonal of $\tilde{\boldsymbol{\it{P}}}^{det}$ and thus compensate for the aberrations in detection.
    \item Convergence: Steps 1-4 are repeated until phase updates converge, yielding the estimated aberration profile, typically within a few tens of iterations on our datasets.
    \item Image reconstruction: Applying the final phase correction to all matrix columns (and rows) reconstructs an aberration free reflection-matrix, where each column represents an aberration free frame. One can use a single frame as the reconstruction or coherently compound the different frames to reconstruct an improved image, which is equivalent to reconstructing a confocal image from structured light illuminations. 
\end{enumerate}

\section{Algorithm implementation}\label{sec_implementation}

\renewcommand{\thefigure}{S1}
\begin{figure*}
\centering{}\includegraphics[width=0.999\textwidth]{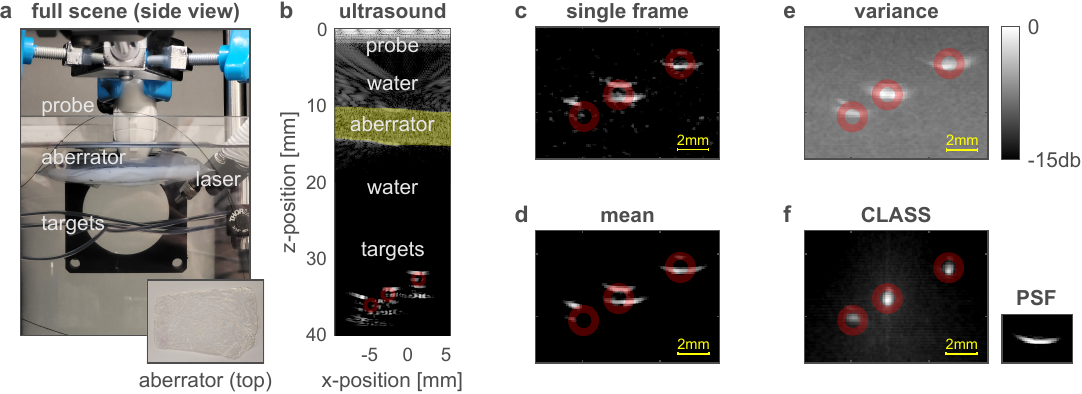}
\caption{\label{fig_experimental_3_circles}
Experimental photoacoustic covariance-matrix imaging through a tissue-mimicking aberrator. 
\textbf{a}, Experimental setup (side view): a linear array ultrasound probe images a vessel like structure composed of three silicon tubes with flowing absorbers (6µm diameter polymer microspheres) through a $\sim$5mm-thick agar-based tissue-mimicking aberrator (inset). 
\textbf{b}, Ultrasound echography image shows the aberrating layer and an aberrated image of the silicon tubes crossing the imaging plane containing flowing absorbing dye solution (tubes walls are marked by red circles in panels b-f). 
\textbf{c}, A single photoacoustic image of the target, displaying the aberrated circular tubes cross-sections. Only top and bottom parts are visible due to the probe limited view.
\textbf{d}, Averaged 440 frames display higher SNR but no clearer features or resolution. 
\textbf{e}, Conventional fluctuation-imaging (auto)variance analysis provides an aberrated view of the flowing absorbers inside the tubes. 
\textbf{f}, Covariance matrix aberration-corrected image, reconstructed by applying the  CLASS algorithm \cite{weinberg_sunray_class_2023, kang_class_2017, lee_class_2022} to the photoacoustic covariance matrix. The computational processing provides both the an improved reconstruction of the flowing absorbers distribution in the true round inner cross-section of the tubes, and the aberrated point spread function (PSF, inset).
Panels b–f share a colorbar shown in panel e.}
\end{figure*}

Key parameters regarding the implementation of the beamforming and reconstruction algorithms are listed below.

\textbf{Memory resources.}
An NVIDIA GeForce RTX 3090 GPU with 24 GB memory was used for running the CLASS algorithm and generating RF data for the numerical investigation. All other calculations were performed on an Intel Core i7-12700 CPU (2.10 GHz) with 32 GB RAM.

\textbf{Data preparation.} For the numerical investigation, RF-data was simulated by the \textit{k-wave} Matlab toolbox, \cite{k_wave_toolbox_2010} as described in Appendix~\ref{subsec_simulating_photoacoustic_data}. Experimental data acquisition is described in Appendix~\ref{subsec_experimental_work}.
Beamforming and CLASS algorithms were applied to an IQ representation of the measured data. For simulated data, the frequency was limited to around 15MHz, maintaining a bandwidth of 80\%. In the case of experimental data, the frequency was restricted to 50\% bandwidth around 7.8 MHz.

\textbf{Beamforming of single frames.}
A standard delay-and-sum beamforming algorithm \cite{PICMUS_2016, Perrot_Garcia_2021, slobodkin_2024} was used to reconstruct each single photoacoustic frame from the measured or simulated signals.

\textbf{Conventional fluctuation imaging.}
Following beamforming of 100 single frames, the temporal variance of each pixel fluctuations was plotted to visualize aberrations in conventional photoacoustic fluctuation imaging. \cite{chaigne_speckle_fluctuation_imaging_2016,chaigne_flow_fluctuation_imaging_2017,vilov_fluctuation_imaging_blood_2020,Godefroy_fluctuation_imaging_3D_2023} To establish a baseline for comparison, an additional dataset was generated without aberrations present, by imaging the same targets in homogeneous media and displaying the corresponding photoacoustic fluctuation images.

\textbf{CLASS implementation.}
Recently, a memory-efficient implementation of the CLASS algorithm was introduced, \cite{weinberg_sunray_class_2023} reducing the   memory requirement to $O(MN)$, where $M$ denotes the number of acquired frames and $N \gg M$ denotes the number of image pixels, rather than the $O(N^2)$ memory size required for the full covariance matrix.
This advancement enables the application of the algorithm to megapixel-scale images without the need to explicitly compute the covariance matrix $\tilde{\boldsymbol{\it{R}}}$. In this work, we used this memory-efficient implementation by running the publicly available Python script from Weinberg et al.\cite{weinberg_sunray_class_2023}

\textbf{Grid sizes and computation times.}
Beamforming with simulated photoacoustic data (Fig.~\ref{fig_simulation_unisoplanatic}): A total of 100 frames, 3562700 pixels each, and a pixel size of 60µm$\times$60µm, were used to represent the full field of view for the conventional fluctuation imaging result in panel b. Beamforming each single frame took 9.9s, corresponding to the reconstruction time of 2.8µs per pixel. Jitter compensation for all frames took 58.7s. Calculating the variance of each pixel fluctuations for 100 frames took 1.2s. Each zoomed-in view (ROI) in panel c consists of 40000 pixels. CLASS reconstruction for each ROI, using 100 frames, 40000 pixels each, and 100 iterations of the algorithm, took 9ms per iteration, with the algorithm typically converging within 20 iterations. 

Beamforming with experimental photoacoustic data (Fig.~\ref{fig_experimental_knot}): The experimental target was depicted using 19881 pixels with a pixel size of 100µm$\times$100µm. Computation time for beamforming a single frame was 0.05s. Jitter compensation for 200 frames took 53s. Mean and variance images in panels e and f were obtained within 15ms. 100 iterations of the CLASS algorithm, using 200 frames, took 1.1s (panels g and h).

\section{Additional experimental results}\label{sec_experimental_additonal_results}

To further validate the robustness of the covariance-matrix approach, we performed an additional experiment, detailed in Fig.~\ref{fig_experimental_3_circles}, where three silicon tubes (1~mm inner diameter, 2~mm outer diameter) are arranged to cross the imaging plane nearly perpendicularly at varying depths.

As with the results presented in the main text, the introduction of a 5~mm-thick agar-based tissue-mimicking aberrator (Fig.~\ref{fig_experimental_3_circles}a,b) significantly distorts the initial pressure reconstruction. Standard averaging of 440 frames (Fig.~\ref{fig_experimental_3_circles}d) improves the signal-to-noise ratio (SNR) but fails to recover the circular cross-section of the tubes, showing only the top and bottom boundaries. Similarly, conventional photoacoustic fluctuation imaging (autovariance, Fig.~\ref{fig_experimental_3_circles}e) remains aberrated.

By applying the CLASS algorithm to the virtual reflection matrix constructed from the temporal fluctuations, we successfully compensate for the distortions induced by the agar layer and the silicone walls of the tubes. The corrected image (Fig.~\ref{fig_experimental_3_circles}f) recovers the circular cross-sections of the inner volume of the tubes. CLASS also provides an estimate of the aberrated point spread function (PSF, Fig.~\ref{fig_experimental_3_circles}f inset).



%
%

%


\bibliography{Library}

\end{document}